\documentclass{IEEEtran}
\usepackage{amssymb}
\usepackage{amsfonts}
\usepackage{amsmath}
\usepackage{algorithm}
\usepackage{tikz}
\usepackage{graphicx,subfigure,cite,multicol,multirow,diagbox,booktabs,array}
\usepackage{color}
\usepackage{stfloats}
\usepackage{subfigure}  
\usepackage{cite}
\usepackage{amsmath,amssymb,amsfonts}
\usepackage{algorithmicx}  
\usepackage{algorithm}
\usepackage{graphicx,color}
\usepackage{textcomp}
\usepackage{graphicx}
\usepackage{epstopdf}
\usepackage{algpseudocode}

\UseRawInputEncoding

\ifCLASSINFOpdf
\else
\fi
\title{\huge Multi-stream Transmission for Directional Modulation Network via Distributed Multi-UAV-aided Multi-active-IRS}
\author{Ke Yang, Rongen Dong, Wei Gao, Feng Shu,  Weiping Shi, Yan Wang, Xuehui Wang and Jiangzhou Wang

\thanks{This work was supported in part by the National Natural Science Foundation of China (Nos.U22A2002, and 62071234), the Hainan Province Science and Technology Special Fund (ZDKJ2021022), the Scientific Research Fund Project of Hainan University under Grant KYQD(ZR)-21008, the Collaborative Innovation Center of Information Technology, Hainan University (XTCX2022XXC07), and the National Key Research and Development Program of China under Grant 2023YFF0612900. \emph{(Corresponding authors: Feng Shu; Rongen Dong; Wei Gao)}}

\thanks{Ke Yang, Rongen Dong, Feng Shu, Yan Wang, and Xuehui Wang are with the School of Information and Communication Engineering, Hainan University, Haikou 570228, China. (Email: shufeng0101@163.com)}

\thanks{Feng Shu is with the School of Electronic and Optical Engineering, Nanjing University of Science and Technology, Nanjing 210094, China. (Email: shufeng0101@163.com)}

\thanks{Weiping Shi is with the School of Network and Communication, Nanjing Vocational College of Information Technology, Nanjing, 210023, China. }

}

\begin{document}

\maketitle

\begin{abstract}
	Active intelligent reflecting surface (IRS) is a revolutionary technique for the future 6th generation mobile networks. The conventional far-field single-IRS-aided directional modulation (DM) networks have only one (no direct path) or two (existing direct path) degrees of freedom (DoFs). This means that there are only one or two streams transmitted simultaneously from base station to user and will seriously limit its rate gain achieved by IRS. How to create multiple DoFs more than two for DM?  In this paper, single large-scale IRS is divided to multiple small IRSs and a novel multi-IRS-aided multi-stream DM network is proposed to achieve a point-to-point multi-stream transmission by creating $K$ ($\geq3$) DoFs, where  multiple small  IRSs are placed distributively via multiple unmanned aerial vehicles (UAVs). The null-space projection, zero-forcing (ZF) and phase alignment are adopted to obtain the transmit beamforming, receive beamforming and phase shift matrix (PSM), respectively, called NSP-ZF-PA. Here, $K$ PSMs and their corresponding  beamforming vectors are independently optimized.  The weighted minimum mean-square error (WMMSE) algorithm is involved in alternating iteration for the optimization variables by introducing the power constraint on IRS, named WMMSE-PC, where the majorization-minimization (MM) algorithm is utilized to address the total PSM. To achieve a lower computational complexity, a maximum trace method, called Max-TR-SVD, is proposed by optimizing the PSM of all IRSs. Simulation results have shown that the proposed NSP-ZF-PA  performs much better than Max-TR-SVD in terms of rate. In particular, the rate of NSP-ZF-PA with sixteen small IRSs is about five times that of NSP-ZF-PA  with combining all small IRSs as a single large IRS. Thus, a dramatic rate enhancement may be achieved by multiple distributed IRSs.

\end{abstract}


\section{INTRODUCTION}
\IEEEPARstart{N}{owadays}, the rapid development of wireless network has greatly improved one's lives. Due to its ultra-high rate and direction of arrival (DOA) measurement precision, the large-scale multi-input multi-output (MIMO) array has already become a research hotspot in academic and industry field \cite{b1, b2, b3}. The deep-learning (DL) transmission design and receiver scheme for physical layer communication network were investigated in \cite{b4}. With the upgrade of meta-materials, passive intelligent reflecting surface (IRS) is getting more and more attention from researchers for its possible and potential applications  in future wireless networks \cite{b5}. For the problems of MIMO channel estimation and signal detection in the model-driven DL network, \cite{b6} developed an orthogonal approximate message passing (OAMP) detection algorithm to solve it,  and validated that the proposed OAMP algorithm performed better than traditional detectors. Passive large IRS is an important physical layer technology of B5G systems, since it can enhance network's quality of service, extend coverage, and reduce  power consumption \cite{b7}. Passive IRS contains a large quantity of affordable reflective elements, it is controlled intelligently, and  can effectively control the wavefront of the incident signals. Passive IRS intelligently configures the wireless environment between base station (BS) and user to improve communication rate \cite{b8, b9, b10}. In \cite{b11}, the authors made an investigation of the design of beamforming methods, the detection of passive IRS channel, the equipment of IRS, and advised that the machine learning technique may be an useful way to make an optimal deployment of IRS. There are many explorations in the passive IRS, which were combined with the practical application. The authors in \cite{b12} designed a prototype of passive IRS-aided wireless network and verified that this prototype could reduce the cost of hardware and power. \cite{b13} introduced the passive IRS to improve the stability of millimeter wave communication, and the co-optimization was utilized to obtain the phase shift matrix (PSM) of IRS and beamforming of transmitter. The energy conservation in passive IRS-aided network has also been investigated in \cite{b14, b15}, the former used the gradient descent and fractional programming to figure out the power distribution and phase shift, while the latter utilized the semidefinite relaxation (SDR) to acquire the PSM at IRS. In \cite{b16}, the authors proposed a novel IRS-aided network to achieve the amplitude-phase modulation by controlling the ON-OFF state and phase shift. The authors in \cite{b17} proposed a method to estimate channel, cyclic-prefix single-carrier cyclic delay diversity with the aid of IRS to alternately transmit the pilots and data, which is robust in time-varying conditions.  

Active IRS can overcome the inherent physical limitation, the effect of double fading introduced by IRS, and further increase the rate gain. It is mainly due to the fact that the active IRS reflects signals with an amplification capability while the passive IRS just adjusts the phase of incident signals and no amplitude. And active IRS can attain more throughput than passive IRS for small-scale or medium-scale scenarios \cite{b18, b19, b20, b21}. Similarly, active IRS can also be combined with other technologies. The alternating iteration is exploited to alternately optimize the beamforming at BS and IRS, and it was shown that the two active IRSs can make a higher rate gain than one or two passive IRSs in \cite{b22}.  The authors in \cite{b23} researched the networks of non-orthogonal multiple access with the help of the active IRS, the achievable rate was promoted by utilizing SDR and minimum-mean-square-error (MMSE). The norm and normalized beamforming vector were alternately iterated in \cite{b24} to acquire the optimization approximation signal-to-noise ratio (SNR). The authors in \cite{b25} used the linear minimum mean square error to estimate the cascaded channel in the active IRS-aided network.\cite{b26} employed average SNR at active IRS to strengthen the capability of network and revealed that the SNR at receiver is proportional to the SNR at IRS. The authors in \cite{b27} derived the security and dependability in an active IRS-aided network by analyzing the outage and intercept probability, and it was verified that active IRS performs better than passive IRS.

Directional modulation (DM) technique can enhance the pattern of antenna array to the target direction and weaken  it in unexpected directions. And it is widely used in physical-layer security  field, which have been investigated  in \cite{b28, b29, b30}. The authors in \cite{b31} utilized the artificial noise (AN) projection and phase alignment to transmit signals securely and accurately in DM network. The IRS can also be combined with DM networks to overcome the limitation that traditional DM network can only transmit one privacy data stream. The passive IRS in \cite{b32} was used to create more than one transmit path, the null-space projection and alternating iterative algorithm were utilized to promote the secrecy rate (SR). In \cite{b33}, to enhance the communication security in the DM network with passive IRS, the PSM and receive beamforming were optimized by alternating iteration, and the zero forcing (ZF) method was also been applied to reduce complexity. To raise the performance of the DM network with a passive IRS, \cite{b34} utilized AN and ZF to optimize the beamforming in DM network with an active IRS and achieved obvious SR enhancement than no IRS. 

For a far-field single IRS-aided DM network, there is
only one (no direct path) or two (existing direct path) degrees of freedom (DoFs). This will seriously limit its rate gain achieved by IRS. According to the basic principle of MIMO network, DoF plays a prominent role in rate improvement. In this paper, a large-scale active IRS is partitioned into $K$ small distributed IRSs to create a large number of DoFs ($\geq 3$), where each UAV hangs one small IRS. 
This will lead to the maximum DoF of the total network being  $K$ or $K+1$. In other words, in such a  DM network, a multi-stream point-to-point transmission may be achieved over single IRS-aided DM network to make a significant rate enhancement. Our main contributions in this paper can be concluded as follows: 

\begin{enumerate}
	
	\item To achieve a significant rate improvement, a novel multi-IRS-aided multi-stream DM network is proposed. In other words, a large-scale IRS  should  be splitted into several smaller IRSs and these IRSs should be  distributedly placed in free space to create more DoF. This new DM network may implement a point-to-point multi-stream transmission by creating more DoFs than traditional single-IRS DM network  due to a distributed multi-IRS network with each UAV hanging one IRS. When the number of small IRSs is $K$, the achievable  maximum DoF  is $K$ or $K+1$, where $K$ means the direct transmission link from BS to user is blocked and  $K+1$ means the direct communication link from BS to user exists. In this paper, we will focus the former in order to simplify the design of beamforming. However, the beamforming methods may be extended to the latter with some modifications.
	
	\item To eliminate the interference among different data-streams associated with small IRSs, the null-space projection (NSP) at BS and receive zero-forcing (ZF) at user is adopted to design the beamforming of transmitter and receiver. This will lead to the fact that $K$ streams are independently transmitted along the corresponding small IRSs. This method is called NSP-ZF, and has a closed-form and thus low-complexity. Phase alignment method is utilized to 
	calculate the PSM of each small IRSs. We call this method as NSP-ZF-PA, where PA is short for phase alignment. And simulation results have shown that the proposed NSP-ZF-PA  performs much better than proposed Max-TR-SVD in terms of rate. In particular, the rate of NSP-ZF-PA with sixteen sub-IRSs is about five times that of NSP-ZF-PA  with all small IRSs combining a single large IRS.

	\item In the above method, the PSMs of all IRSs are individually optimized. Now, all small IRSs are viewed as a large virtual IRS, the  weighted minimum mean-square error (WMMSE) is proposed with power constraint (PC) on active IRS. Due to the determinant, the original optimization problem is rewritten as a more tractable form. The alternating iteration is utilized to acquire the beamforming at transmitter, PSM and beamforming at receiver. The majorization-minimization (MM) algorithm is involved to obtain the optimization PSM of this large-scale IRS. This method is called WMMSE-PC. To achieve a low computational complexity, the transmit and receive beamforming are designed by singular value decomposition (SVD). Then, a maximum trace method is proposed to optimize the total PSM of all sub-IRSs. Due to its quadratic form, its closed form is directly derived. This methods is called Max-TR-SVD. According to simulation result, we found that the WMMSE-PC has a better performance than NSP-ZF-PA when the power at IRS is low, while having a worse performance than NSP-ZF-PA when the power at IRS is high. And the proposed WMMSE-PC performs also better than Max-TR-SVD about rate.

\end{enumerate}

The remainder is mainly organized as follows. The system model and DoF analysis are shown in Section \uppercase\expandafter{\romannumeral2}. Section \uppercase\expandafter{\romannumeral3} denotes the proposed three methods. The performance and complexity analysis are shown in \uppercase\expandafter{\romannumeral4}. Simulation results we obtained are presented in \uppercase\expandafter{\romannumeral5}. Finally, conclusions are provided in \uppercase\expandafter{\romannumeral6}. 

\textit{Notations:} The sign $\mathbb{C}$ presents the set of complex. Signs $[\cdot]^{\ast}$, $[\cdot]^{\dagger}$, $[\cdot]^{T}$, $[\cdot]^{-1}$ and $[\cdot]^{H}$ express the conjugate, pseudo-inverse, transpose, inverse and conjugate-transpose operations, respectively. The notations $\|\cdot\|$ and $\|\cdot\|_F$ stand for the 2-norm and F-norm operations, respectively. The $\text{diag}(\cdot)$ signifies the diagonal operator. The $\mathbf{I}_N$ is an $N\times N$ unit matrix. The $Tr(\cdot)$ and $|\cdot|$ denote the trace and determinant operation. The notations $Re\{s\}$ and $arg(s)$ indicate the real part and the argument of the value $s$. Signs $\mathbb{E}\{\cdot\}$, $\odot$ and $ln(\cdot)$ express the expectation, Hadamard product and  natural logarithm operators, respectively. The sign $\mathcal{CN}\left({\mathbf{\mu},\sigma^2}\right)$ signifies the complex Gaussian distribution with mean $\mathbf{\mu}$ and variance $\sigma^2$.

\section{SYSTEM MODEL AND DoF ANALYSIS}
\subsection{PARTITIONING A LARGE-SCALE ACTIVE IRS  INTO $K$ SMALLER IRSs}
\begin{figure}[htb]
	\centering
	\subfigure[\scriptsize Splitting a large-scale active IRS into $K$ smaller IRSs.]{
		\label{fig1_a} 
		\includegraphics[width=3.5in]{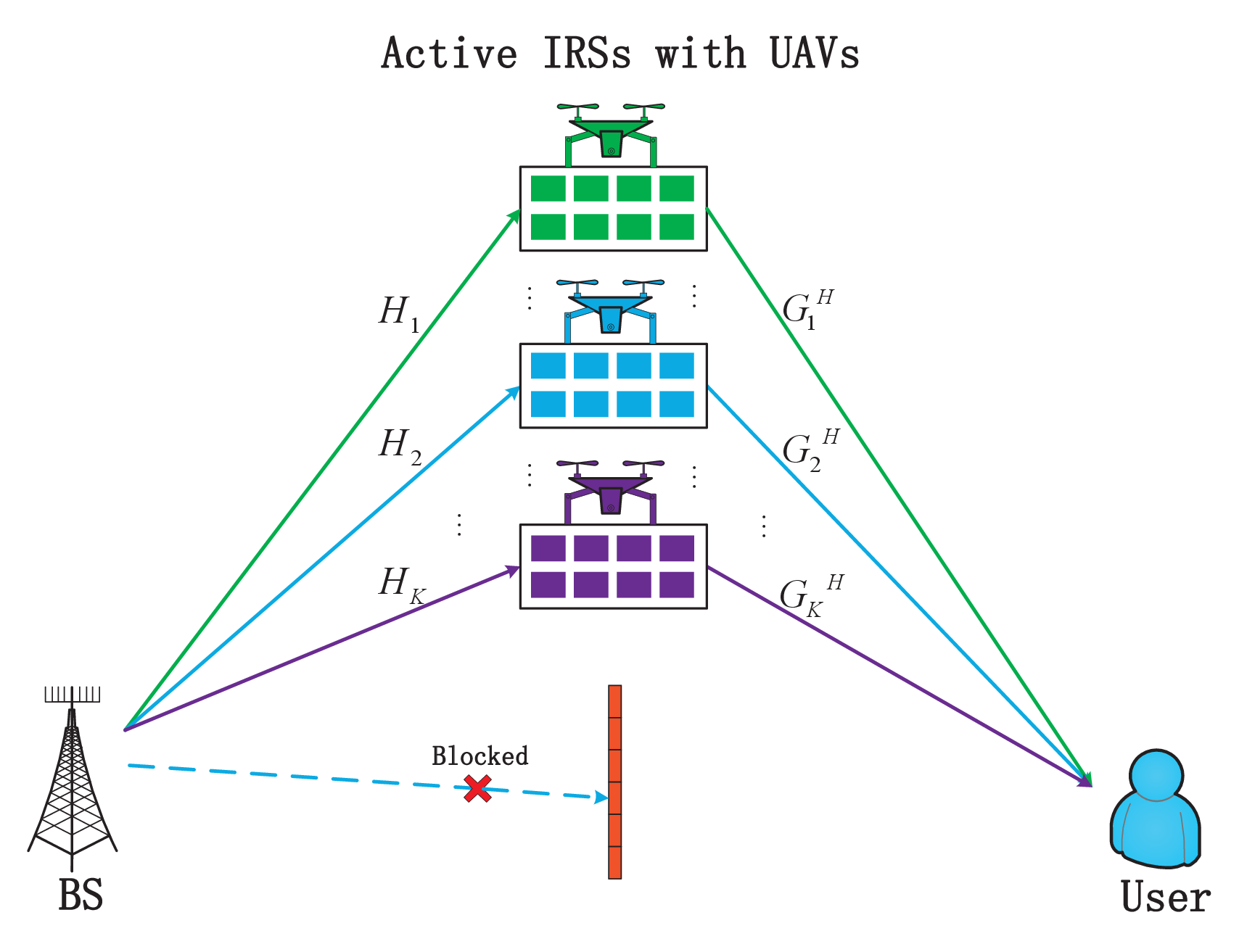}} 
	\hspace{20pt} 
	\subfigure[\scriptsize Viewing $K$ smaller IRSs as a large virtual IRS.]{
		\label{fig1_b} 
		\includegraphics[width=3.5in]{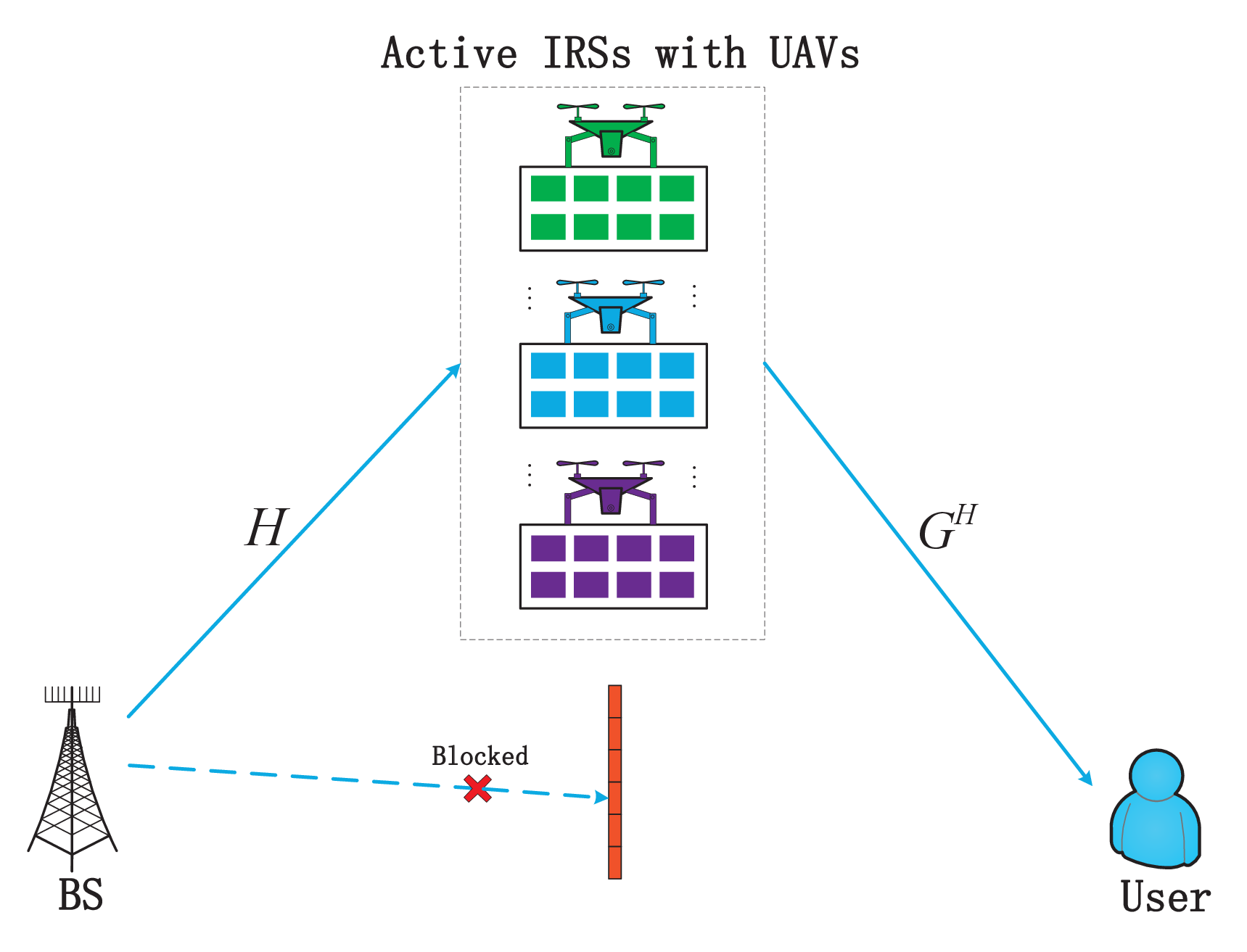}}
	\hspace{20pt}
	\caption{System model of a distributed multi-UAV-aided multi-active-IRS network.}
	\label{fig1} 
\end{figure}

In Fig. \ref{fig1_a}, we plot a distributed multi-UAV-aided multi-active-IRS wireless communication network and a large-scale active IRS is split into $K$ smaller distributed IRSs. There are $M$ antennas in BS, where $M\geq K/K+1$ and the user is equipped with $N_u$ antennas, where $N_u\geq K/K+1$. Besides, there are equipped with $K$ small IRSs, each with $N_k$ reflective elements. Let us define $N_I=KN_k$, and $N_I$ is the sum of numbers of all smaller IRS elements. The line-of-propagation channels are assumed in this paper.

The transmit baseband signal is
\begin{equation}
	\mathbf{x}=\displaystyle \sum_{i=1}^{K}\mathbf{v}_is_i,
\end{equation}
where $\mathbf{v}_i\in\mathbb{C}^{M\times 1}$ is the $i$-th beamforming at BS with $\|\mathbf{v}_i\|^2=1$, and $s_i$ is the $i$-th transmission symbol.

The reflect signal of the $k$-th IRS is
\begin{equation}
	\mathbf{y}_{k}^t=\mathbf{\Theta}_k\mathbf{H}_{k}\mathbf{x}+\mathbf{\Theta}_k\mathbf{n}_{k},
\end{equation}
where $\mathbf{H}_{k}\in\mathbb{C}^{N_k\times M}$ denotes the transmission channel from the BS to the $k$-th IRS, and $\mathbf{\Theta}_k=\text{diag}(\alpha_{1} e^{j\theta_1}, \cdots, \alpha_{N_k} e^{j\theta_{N_k}}) \in\mathbb{C}^{N_k\times N_k}$  stands for the reflection coefficient matrix of $k$-th IRS, where $\alpha_{l}$ represents amplitude gain and $\theta_{l}$ stands for phase shift, respectively. And $\mathbf{n}_{k} \in\mathbb{C}^{N_k\times 1}$ is the additive white Gaussian noise  
(AWGN) at IRS $k$, and $\mathbf{n}_{k}\sim \mathcal{CN}\left({0,{~\sigma^{2}_{k}\mathbf{I}_{N_k}}}\right)$.

The $k$-th signal at user is represented as
\begin{align}
	\mathbf{y}_k
	&=\mathbf{G}_k^H\mathbf{\Theta}_k\mathbf{H}_{k}\mathbf{x}+\mathbf{G}_k^H\mathbf{\Theta}_k\mathbf{n}_{k},
\end{align}
where $\mathbf{G}_k^H\in\mathbb{C}^{N_u\times N_k}$ denotes the channel from IRS $k$ to user.

The total signal received at user is
\begin{align}
	\mathbf{y}_r
	&=\displaystyle \sum_{j=1}^{K}\mathbf{G}_j^H\mathbf{\Theta}_j\mathbf{H}_{j}\mathbf{x}+\displaystyle \sum_{j=1}^{K}\mathbf{G}_j^H\mathbf{\Theta}_j\mathbf{n}_{j}+\mathbf{z}\nonumber\\
	&=\displaystyle \sum_{j=1}^{K}\displaystyle \sum_{i=1}^{K}\mathbf{G}_j^H\mathbf{\Theta}_j\mathbf{H}_{j}\mathbf{v}_is_i+\displaystyle \sum_{j=1}^{K}\mathbf{G}_j^H\mathbf{\Theta}_j\mathbf{n}_{j}+\mathbf{z}.
\end{align}

The $k$-th data-stream signal of user is
\begin{align}
	y_k\label{y}
	&=\mathbf{u}_k^H\mathbf{y}_r=\displaystyle \sum_{j=1}^{K}\displaystyle \sum_{i=1}^{K}\mathbf{u}_k^H\mathbf{G}_j^H\mathbf{\Theta}_j\mathbf{H}_{j}\mathbf{v}_is_i\notag\\&+\displaystyle \sum_{j=1}^{K}\mathbf{u}_k^H\mathbf{G}_j^H\mathbf{\Theta}_j\mathbf{n}_{j}+\mathbf{u}_k^H\mathbf{z},
\end{align}
where $\mathbf{u}_k\in\mathbb{C}^{N_u\times 1}$ denotes the $k$-th receive beamforming vector with $\|\mathbf{u}_k\|^2=1$, and $\mathbf{z} \in\mathbb{C}^{N_u\times 1}$ is the AWGN at user, where $\mathbf{z}\sim \mathcal{CN}\left({0,{~\sigma^{2}_{z}\mathbf{I}_{N_u}}}\right)$.

\subsection{VIEWING $K$ SMALLER IRSs AS A LARGE VIRTUAL IRS}

In Fig. \ref{fig1_b}, we view all of the distributed smaller IRSs as a virtual large-scale IRS. Here, the  BS and user are still employed with $M$ and $N_u$ antennas, respectively.

Similarly, the total signal received at user is
\begin{align}
	\mathbf{y}_r
	&=\mathbf{G}^H\mathbf{\Theta}\mathbf{H}\mathbf{V}\mathbf{s}+\mathbf{G}^H\mathbf{\Theta}\mathbf{n}+\mathbf{z},
\end{align}
where $\mathbf{G}^H\in\mathbb{C}^{N_u\times N_I}$ , $\mathbf{\Theta}\in\mathbb{C}^{N_I\times N_I}$ and $\mathbf{H}\in\mathbb{C}^{N_I\times M}$ respectively stand for the channel between this large-scale IRS with user, the PSM of IRS and the transmission channel from BS to IRS.  $\mathbf{V}\in\mathbb{C}^{M\times K}$ denotes the transmit beamforming matrix. $\mathbf{s}\in\mathbb{C}^{K\times 1}$ is the transmission symbol vector.   $\mathbf{n}\in\mathbb{C}^{N_I\times 1}$ is the AWGN of this large-scale IRS with distribution $\mathbf{n}\sim \mathcal{CN}\left({0,{~\sigma^{2}_{n}\mathbf{I}_{N_I}}}\right)$, where $\sigma^{2}_{n}=K\sigma^{2}_{k}$.

Then, the total receive signal at user is rewritten as
\begin{align}
	\mathbf{y}\label{y'}
	&=\mathbf{U}^H\mathbf{G}^H\mathbf{\Theta}\mathbf{H}\mathbf{V}\mathbf{s}+\mathbf{U}^H\mathbf{G}^H\mathbf{\Theta}\mathbf{n}+\mathbf{U}^H\mathbf{z},
\end{align}
where $\mathbf{U}^H\in\mathbb{C}^{K\times N_u}$ denotes the receive beamforming matrix.

	\subsection{DoF ANALYSIS}
	Since there are $K$ IRSs in the proposed multi-active IRS-aided DM network, a maximum of $K$ DoFs is created. The specific derivation process is as follows
	\begin{align}
		\text{rank}(\mathbf{H})
		&\leq \min(M, K),
	\end{align}
	\begin{align}
		\text{rank}(\mathbf{G})
		&\leq \min(K, N_u),
	\end{align}
	where 
	\begin{align}
		\mathbf{H}=[\mathbf{H}_1^H, \mathbf{H}_{2}^H, \cdots,\mathbf{H}_k^H,\cdots,\mathbf{H}_K^H]^H,
	\end{align}
	and
	\begin{align}
		\mathbf{G}=[\mathbf{G}_1^H, \mathbf{G}_{2}^H,\dots,\mathbf{G}_k^H,\dots,\mathbf{G}_K^H]^H.
	\end{align}
	Then the DoF of our network can be written as
	\begin{align}\label{dof1}
		\text{DoF}
		&\leq \min(\text{rank}(\mathbf{H}), \text{rank}(\mathbf{G}))=\min(K, M, N_u).
	\end{align}
	Based on (\ref{dof1}), the maximum DoF of our network is under the condition that all $K$ smaller IRSs are randomly placed such that the $K$ transmit steering vectors  departing from BS  are linear independence and form a basis of  transmit signal space. Similarly, the $K$ receive steering vectors impinging on receive array at user are also linear independence and forms a basis of the receive signal space. Thus, $K$ parallel independent bit streams may be delivered from BS to user and it is necessary for us to set the $M\geq K/K+1$ and $N_u\geq K/K+1$. In this case, the maximum DoF of the system is $K$. Then the BS can transmit $K$ data-streams simultaneously.

\section{PROPOSED THREE METHODS}
\subsection{PROPOSED NSP-ZF-PA}
To reduce the interference from other data-stream to the $k$-th data-stream signal, we project other stream interference information onto $k$-th null-space of transmission channel.

Define 
$\mathbf{H}_k\mathbf{v}_{j}=\mathbf{0}$, where $k\neq j$, then the $k$-th data-stream signal of user in (\ref{y}) can be rewritten as
\begin{align}\label{yk}
	y_k
	&=\mathbf{u}_k^H\mathbf{G}_k^H\mathbf{\Theta}_k\mathbf{H}_{k}\mathbf{v}_ks_k+\displaystyle \sum_{j=1,j\neq k}^{K}\mathbf{u}_k^H\mathbf{G}_j^H\mathbf{\Theta}_j\mathbf{H}_{j}\mathbf{v}_js_j\notag\\&+\displaystyle \sum_{j=1}^{K}\mathbf{u}_k^H\mathbf{G}_j^H\mathbf{\Theta}_j\mathbf{n}_{j}+\mathbf{u}_k^H\mathbf{z},
\end{align}
based on zero-forcing theorem, let   
$\mathbf{u}_i^H\mathbf{G}_k^H\rightarrow0$, where $i\neq k$, then the power of $k$-th data-stream at user is given by
\begin{align}
	\mathbb{E}\{y_k^Hy_k\}
	&=|\mathbf{u}_k^H\mathbf{G}_k^H\mathbf{\Theta}_k\mathbf{H}_{k}\mathbf{v}_ks_k|^2+\displaystyle \sum_{j=1}^{K}\sigma_j^2\|\mathbf{u}_k^H\mathbf{G}_j^H\mathbf{\Theta}_j\|^2\notag\\&+\displaystyle\sum_{j=1,j\neq k}^{K}|\mathbf{u}_k^H\mathbf{G}_j^H\mathbf{\Theta}_j\mathbf{H}_{j}\mathbf{v}_js_j|^2+\sigma_z^2\|\mathbf{u}_k^H\|^2.
\end{align}
The SINR of the $k$-th data-stream is
\begin{align}\label{SINR}
	\gamma_k
	&=\frac{|\mathbf{u}_k^H\mathbf{G}_k^H\mathbf{\Theta}_k\mathbf{H}_{k}\mathbf{v}_ks_k|^2}{A},
\end{align}
where 
\begin{align}
	A
	&=\displaystyle\sum_{j=1,j\neq k}^{K}|\mathbf{u}_k^H\mathbf{G}_j^H\mathbf{\Theta}_j\mathbf{H}_{j}\mathbf{v}_js_j|^2+\displaystyle \sum_{j=1}^{K}\sigma_j^2\|\mathbf{u}_k^H\mathbf{G}_j^H\mathbf{\Theta}_j\|^2\notag\\&+\sigma_z^2\|\mathbf{u}_k^H\|^2.
\end{align}

As thus, the rate of user is written as
\begin{align}\label{Rm1}
	R_{m_1}=\displaystyle\sum_{k=1}^{K}\log_2(1+\gamma_k).
\end{align}

Next, let's focus on the specific design of transmit beamforming vector $\mathbf{v}_k$, receive beamforming vector $\mathbf{u}_k$ and the PSM $\mathbf{\Theta}_k$. 

Let us define $\mathbf{H}_{-k}=[\mathbf{H}_1^H,\dots,\mathbf{H}_{k-1}^H,\mathbf{H}_{k+1}^H,\dots,\mathbf{H}_K^H]^H$, and fix $\mathbf{u}_k, \mathbf{\Theta}_k$, then the optimization problem with respect to $\mathbf{v}_k$ is presented as

\begin{align}\label{vk}
	&\max_{\mathbf{v}_k}~~~~	\mathbf{v}_k^H\mathbf{H}_k^H\mathbf{H}_k\mathbf{v}_k\nonumber\\
	&~~\text{s.t.}~~~~
	\mathbf{H}_{-k}\mathbf{v}_k=\mathbf{0},\nonumber\\
	&~~~~~~~~~~\mathbf{v}_k^H\mathbf{v}_k=1. 		
\end{align}

Define $\mathbf{T}_{-k}=[\mathbf{I}_M-\mathbf{H}_{-k}^H(\mathbf{H}_{-k}\mathbf{H}_{-k}^H)^{\dagger}\mathbf{H}_{-k}]$, and $\mathbf{v}_k=\mathbf{T}_{-k}\boldsymbol{\alpha}_k$, (\ref{vk}) can be converted as

\begin{align}
	&\max_{\boldsymbol{\alpha}_k}~~~~	\boldsymbol{\alpha}_k^H\mathbf{T}_{-k}^H\mathbf{H}_k^H\mathbf{H}_k\mathbf{T}_{-k}\boldsymbol{\alpha}_k\nonumber\\
	&~~\text{s.t.}~~~~~~\boldsymbol{\alpha}_k^H\boldsymbol{\alpha}_k=1. 		
\end{align}

Define $\mathbf{B}_k=\mathbf{T}_{-k}^H\mathbf{H}_k^H\mathbf{H}_k\mathbf{T}_{-k}$, noted that $\mathbf{B}_k$ is a Hermitian matrix, then it can be decomposed as
\begin{align}
	\mathbf{B}_k
	&=\mathbf{E}_k\mathbf{\Lambda}_1 \mathbf{E}_k^H=\displaystyle\sum_{i=1}^{M}\lambda_i\mathbf{e}_i\mathbf{e}_i^H, 		
\end{align}
where $\mathbf{\Lambda}_1=\text{diag}(\lambda_1,\lambda_2,\dots,\lambda_M)$, with $\lambda_1\leq\lambda_2\leq\dots\leq\lambda_M$, according to the Rayleigh-Ritz theorem,  we have $\boldsymbol{\alpha}_k=\mathbf{e}_M$.

Then the transmit beamforming vector is
\begin{align}
	\mathbf{v}_k
	&=\frac{\mathbf{T}_{-k}\mathbf{e}_M}{\|\mathbf{T}_{-k}\mathbf{e}_M\|}. 		
\end{align}

Similarly, let $\mathbf{G}_{-k}=[\mathbf{G}_1^H,\dots,\mathbf{G}_{k-1}^H,\mathbf{G}_{k+1}^H,\dots,\mathbf{G}_K^H]$, and fix $\mathbf{v}_k, \mathbf{\Theta}_k$, then the optimization problem with respect to $\mathbf{u}_k$ is
\begin{align}\label{uk}
	&\max_{\mathbf{u}_k}~~~~	\mathbf{u}_k^H\mathbf{G}_k^H\mathbf{G}_k\mathbf{u}_k\nonumber\\
	&~~\text{s.t.}~~~~~
	\mathbf{u}_k^H\mathbf{G}_{-k}=\mathbf{0},\nonumber\\
	&~~~~~~~~~~~\mathbf{u}_k^H\mathbf{u}_k=1. 		
\end{align}

Define $\mathbf{L}_{-k}=[\mathbf{I}_{N_u}-\mathbf{G}_{-k}(\mathbf{G}_{-k}^H\mathbf{G}_{-k})^{\dagger}\mathbf{G}_{-k}^H]$, and $\mathbf{u}_k=\mathbf{L}_{-k}\boldsymbol{\zeta}_k$, (\ref{uk}) can be rewritten as 
\begin{align}
	&\max_{\boldsymbol{\zeta}_k}~~~~	\boldsymbol{\zeta}_k^H\mathbf{L}_{-k}^H\mathbf{G}_k^H\mathbf{G}_k\mathbf{L}_{-k}\boldsymbol{\zeta}_k\nonumber\\
	&~~\text{s.t.}~~~~~~\boldsymbol{\zeta}_k^H\boldsymbol{\zeta}_k=1. 		
\end{align}

Let $\mathbf{C}_k=\mathbf{L}_{-k}^H\mathbf{G}_k^H\mathbf{G}_k\mathbf{L}_{-k}$, noted that $\mathbf{C}_k$ is also a Hermitian matrix, then it can be decomposed as
\begin{align}
	\mathbf{C}_k
	&=\mathbf{F}_k\mathbf{\Lambda}_2 \mathbf{F}_k^H=\displaystyle\sum_{i=1}^{N_u}\mu_i\mathbf{f}_i\mathbf{f}_i^H, 		
\end{align}
where $\mathbf{\Lambda}_2=\text{diag}(\mu_1,\mu_2,\dots,\mu_{N_u})$, and $\mu_1\leq\mu_2\leq\dots\leq\mu_{N_u}$, according to the Rayleigh-Ritz theorem, we have $\boldsymbol{\zeta}_k=\mathbf{f}_{N_u}$.

Then the receive beamforming vector is shown as
\begin{align}
	\mathbf{u}_k
	&=\frac{\mathbf{L}_{-k}\mathbf{f}_{N_u}}{\|\mathbf{L}_{-k}\mathbf{f}_{N_u}\|}. 		
\end{align}

The $k$-th useful signal power at user is
\begin{align}
	P_k=\mathbf{v}_k^H\mathbf{H}_{k}^H\mathbf{\Theta}_k^H\mathbf{G}_k\mathbf{u}_k\mathbf{u}_k^H\mathbf{G}_k^H\mathbf{\Theta}_k\mathbf{H}_{k}\mathbf{v}_k.
\end{align}

Let $\mathbf{\Theta}_k=\text{diag}(\boldsymbol{\theta}_k)$, fixed $\mathbf{v}_k$ and $\mathbf{u}_k$, and define $\boldsymbol{\theta}_k=\tilde{\boldsymbol{\theta}}_k\tilde{\rho}_k$, where $\tilde{\rho}_k=\|\boldsymbol{\theta}_k\|$, $\tilde{\boldsymbol{\theta}}_k^H\tilde{\boldsymbol{\theta}}_k=1$. Based on the phase alignment (PA) theorem, the $k$-th useful signal power at user can be represented as
\begin{align}
	P_k
	&=\boldsymbol{\theta}_k^H\text{diag}(\mathbf{H}_k\mathbf{v}_k)^H\mathbf{G}_k\mathbf{u}_k\mathbf{u}_k^H\mathbf{G}_k^H\text{diag}(\mathbf{H}_k\mathbf{v}_k)\boldsymbol{\theta}_k\notag\\&=\tilde{\rho}_k^2\tilde{\boldsymbol{\theta}}_k^H\text{diag}(\mathbf{H}_k\mathbf{v}_k)^H\mathbf{G}_k\mathbf{u}_k\mathbf{u}_k^H\mathbf{G}_k^H\text{diag}(\mathbf{H}_k\mathbf{v}_k)\tilde{\boldsymbol{\theta}}_k.
\end{align}

Then the $\boldsymbol{\theta}_k$ can be obtained by
\begin{align}
	\boldsymbol{\theta}_k=\tilde{\boldsymbol{\theta}}_k\tilde{\rho}_k=\frac{\mathbf{u}_k^H\mathbf{G}_k^H\text{diag}(\mathbf{H}_k\mathbf{v}_k)}{\|\mathbf{u}_k^H\mathbf{G}_k^H\text{diag}(\mathbf{H}_k\mathbf{v}_k)\|}\tilde{\rho}_k.
\end{align}   

The power reflected by the $k$-th IRS is 
\begin{equation}
	P_{I_k}=\tilde{\rho}_k^2(\|\text{diag}(\mathbf{H}_{k}\mathbf{v}_k)	\tilde{\boldsymbol{\theta}}_k^H\|^2+\sigma_k^2),
\end{equation}
where $P_{I_k}$ denotes the power budget at $k$ small IRS, and
\begin{align}
	\tilde{\rho}_k
	&=\sqrt{\frac{P_{I_k}}{\|\text{diag}(\mathbf{H}_{k}\mathbf{v}_k)	\tilde{\boldsymbol{\theta}}_k^H\|^2+\sigma_k^2}}.
\end{align}

\subsection{PROPOSED WMMSE-PC}
According to (\ref{y'}), the power of received signal at user is
\begin{align}
	\mathbb{E}\{\mathbf{y}\mathbf{y}^H\}\label{Eyy^H}
	&=\mathbf{U}^H\mathbf{G}^H\mathbf{\Theta}\mathbf{H}\mathbf{V}\mathbf{s}\mathbf{s}^H\mathbf{V}^H\mathbf{H}^H\mathbf{\Theta}^H\mathbf{G}\mathbf{U}\notag\\&+\sigma^{2}_{n}\mathbf{U}^H\mathbf{G}^H\mathbf{\Theta}\mathbf{\Theta}^H\mathbf{G}\mathbf{U}+\sigma^{2}_{z}\mathbf{U}^H\mathbf{U}.
\end{align}

As thus, the final rate of user is
\begin{align}\label{hesulv}
	R_{m2}
	&=\log_2|\tilde{\mathbf{B}}^{-1}(\tilde{\mathbf{A}}+\tilde{\mathbf{B}})|\notag\\
	&=\log_2|\mathbf{I}_{K}+\frac{\tilde{\mathbf{A}}}{\tilde{\mathbf{B}}}|,
\end{align}
where $\tilde{\mathbf{A}}=\mathbf{U}^H\mathbf{G}^H\mathbf{\Theta}\mathbf{H}\mathbf{V}\mathbf{s}\mathbf{s}^H\mathbf{V}^H\mathbf{H}^H\mathbf{\Theta}^H\mathbf{G}\mathbf{U}$, and $\tilde{\mathbf{B}}=\sigma^{2}_{n}\mathbf{U}^H\mathbf{G}^H\mathbf{\Theta}\mathbf{\Theta}^H\mathbf{G}\mathbf{U}+\sigma^{2}_{z}\mathbf{U}^H\mathbf{U}$.

Define $\mathbb{E}[\mathbf{s}\mathbf{s}^H]=\mathbf{I}_{M}$, and the total optimization problem can be expressed as
\begin{align}\label{the optimization problem}
	&\max_{\mathbf{\Theta},\mathbf{V}}~~\log_2|\mathbf{I}_{N_u}+\frac{\mathbf{G}^H\mathbf{\Theta}\mathbf{H}\mathbf{V}\mathbf{V}^H\mathbf{H}^H\mathbf{\Theta}^H\mathbf{G}}{\sigma^{2}_{n}\mathbf{G}^H\mathbf{\Theta}\mathbf{\Theta}^H\mathbf{G}+\sigma^{2}_{z}\mathbf{I}_{N_u}}|\nonumber\\
	&~~~\text{s.t.}~~~~~\|\mathbf{V}\|_F^2\leq P_B,\nonumber\\
	&~~~~~~~~~~~\|\mathbf{\Theta}\mathbf{H}\mathbf{V}\mathbf{s}\|^2+\|\mathbf{\Theta}\|_F^2\sigma_n^2\leq P_I.		
\end{align}
where $P_B$ and $P_I$ stand for the power budgets of BS and large-scale IRS, respectively.

Then, the mean-square error (MSE) of user is
\begin{align}
	\mathbf{E}
	&=\mathbb{E}[(\mathbf{y}-\mathbf{s})(\mathbf{y}-\mathbf{s})^H]\notag\\
	&=(\mathbf{U}^H\mathbf{G}^H\mathbf{\Theta}\mathbf{H}\mathbf{V}-\mathbf{I}_K)(\mathbf{U}^H\mathbf{G}^H\mathbf{\Theta}\mathbf{H}\mathbf{V}-\mathbf{I}_K)^H\notag\\&
	+\sigma^{2}_{n}\mathbf{U}^H\mathbf{G}^H\mathbf{\Theta}\mathbf{\Theta}^H\mathbf{G}\mathbf{U}+\sigma^{2}_{z}\mathbf{U}^H\mathbf{U},
\end{align}
then (\ref{the optimization problem}) is converted as 
\begin{align}
	&\max_{\mathbf{W},\mathbf{U},\mathbf{\Theta},\mathbf{V}}~~h(\mathbf{W},\mathbf{U},\mathbf{\Theta},\mathbf{V})\nonumber\\
	&~~~~\text{s.t.}~~~~~\|\mathbf{V}\|_F^2\leq P_B,\nonumber\\
	&~~~~~~~~~~~~\|\mathbf{\Theta}\mathbf{H}\mathbf{V}\mathbf{s}\|^2+\|\mathbf{\Theta}\|_F^2\sigma_n^2\leq P_I,		
\end{align}
where
\begin{align}
	h(\mathbf{W},\mathbf{U},\mathbf{\Theta},\mathbf{V})
	&=\log_2|\mathbf{W}|-Tr(\mathbf{W}\mathbf{E})+K,
\end{align}
where $\mathbf{W}$ is an auxiliary matrix. Now, the new problem is in a more tractable form \cite{b35}.

Then given the first-order derivative function of $h(\mathbf{W},\mathbf{U},\mathbf{\Theta},\mathbf{V})$ with respect to $\mathbf{U}$ and make it equal zero, then the optimal $\mathbf{U}_{opt}$ is given by
\begin{align}\label{U_opt}
	\mathbf{U}_{opt}
	&=(\mathbf{G}^H\mathbf{\Theta}\mathbf{H}\mathbf{V}\mathbf{V}^H\mathbf{H}^H\mathbf{\Theta}^H\mathbf{G}+\tilde{\mathbf{C}})^{-1}\mathbf{G}^H\mathbf{\Theta}\mathbf{H}\mathbf{V},
\end{align}
where $\tilde{\mathbf{C}}=\sigma^{2}_{n}\mathbf{G}^H\mathbf{\Theta}\mathbf{\Theta}^H\mathbf{G}+\sigma^{2}_{z}\mathbf{I}_{N_u}$.

Given $\mathbf{U},\mathbf{\Theta}$ and $\mathbf{V}$, the optimal $\mathbf{W}$ is obtained as
\begin{align}\label{W}
	\mathbf{W}_{opt}
	&=\mathbf{E}^{-1}.
\end{align}

Now, let us focus on the optimizing of the $\mathbf{V}$ and $\mathbf{\Theta}$.

The  optimization problem ($\mathbf{W},\mathbf{U}$ and $\mathbf{\Theta}$ are fixed) with respect to $\mathbf{V}$ is given by
\begin{align}\label{V_opt}
	&\min_{\mathbf{V}}~~Tr(\mathbf{W}\mathbf{E})\nonumber\\
	&~\text{s.t.}~~~~\|\mathbf{V}\|_F^2\leq P_B,\nonumber\\
	&~~~~~~~~\|\mathbf{\Theta}\mathbf{H}\mathbf{V}\mathbf{s}\|^2+\|\mathbf{\Theta}\|_F^2\sigma_n^2\leq P_I.
\end{align}
It can be addressed by utilizing CVX packges, because (\ref{V_opt}) is a convex optimization problem.

The optimization problem ($\mathbf{W},\mathbf{U}$ and $\mathbf{V}$ are fixed) with respect to $\mathbf{\Theta}$ is shown as
\begin{align}\label{Theta_opt}
	&\min_{\mathbf{\Theta}}~~Tr(\mathbf{\Theta}^H\hat{\mathbf{A}}\mathbf{\Theta}\hat{\mathbf{B}})+Tr(\mathbf{\Theta}\hat{\mathbf{C}})+Tr(\mathbf{\Theta}^H\hat{\mathbf{C}}^H)\nonumber\\
	&~~~~~~~~	+Tr(\sigma^{2}_{n}\mathbf{\Theta}^H\hat{\mathbf{A}}\mathbf{\Theta})\nonumber\\
	&~\text{s.t.}~~~\|\mathbf{\Theta}\mathbf{H}\mathbf{V}\mathbf{s}\|^2+\|\mathbf{\Theta}\|_F^2\sigma_n^2\leq P_I,
\end{align}
where $\hat{\mathbf{A}}=\mathbf{G}\mathbf{U}\mathbf{W}\mathbf{U}^H\mathbf{G}^H$, $\hat{\mathbf{B}}=\mathbf{H}\mathbf{V}\mathbf{V}^H\mathbf{H}^H$, $\hat{\mathbf{C}}=-\mathbf{H}\mathbf{V}\mathbf{W}\mathbf{U}^H\mathbf{G}^H$. Based on \cite{b35}, the optimization problem (\ref{Theta_opt}) is converted to
\begin{align}\label{optimization problem1}
	&\min_{\boldsymbol{\theta}}~~f(\boldsymbol{\theta})=\boldsymbol{\theta}^H\mathbf{\Omega}\boldsymbol{\theta}+2Re\{\boldsymbol{\theta}^H\mathbf{c}^*\}\nonumber\\
	&~\text{s.t.}~~~\|\mathbf{\Theta}\mathbf{H}\mathbf{V}\mathbf{s}\|^2+\|\mathbf{\Theta}\|_F^2\sigma_n^2\leq P_I,
\end{align}
where $\mathbf{\Theta}=\text{diag}(\boldsymbol{\theta})$, $\mathbf{C}=\text{diag}(\mathbf{c})$ and $\mathbf{\Omega}=\hat{\mathbf{A}}\odot \hat{\mathbf{B}}^T+\sigma_n^2\hat{\mathbf{A}}\odot \mathbf{I}$.

Then the majorization-minimization (MM) algorithm is utilized to figure out (\ref{optimization problem1}).

Noted that
\begin{align}
	\boldsymbol{\theta}^H\mathbf{\Omega}\boldsymbol{\theta}
	&\leq\boldsymbol{\theta}^H\mathbf{X}\boldsymbol{\theta}-2Re\{\boldsymbol{\theta}^H(\mathbf{X}-\mathbf{\Omega})\boldsymbol{\theta}^t\}\notag\\&+(\boldsymbol{\theta}^t)^H(\mathbf{X}-\mathbf{\Omega})\boldsymbol{\theta}^t=\hat{f}(\boldsymbol{\theta}|\boldsymbol{\theta}^t),
\end{align}
where $\boldsymbol{\theta}^t$ is the value of $\boldsymbol{\theta}$ after $t$ iterations, and $\mathbf{X}=\lambda_{max}(\mathbf{\Omega})\mathbf{I}$, $\lambda_{max}$ is the maximum eigenvalue of $\mathbf{\Omega}$, then the upper bound function of $f(\boldsymbol{\theta})$ is
\begin{align}
	g(\boldsymbol{\theta}|\boldsymbol{\theta}^t)
	&=\hat{f}(\boldsymbol{\theta}|\boldsymbol{\theta}^t)+2Re\{\boldsymbol{\theta}^H\mathbf{c}^*\}.
\end{align}

Let us define $P_I=\|\mathbf{\Theta}\mathbf{H}\mathbf{V}\mathbf{s}\|^2+\|\mathbf{\Theta}\|_F^2\sigma_n^2$, and $\boldsymbol{\theta}=\gamma\tilde{\boldsymbol{\theta}}$, where $\tilde{\boldsymbol{\theta}}^H\tilde{\boldsymbol{\theta}}=KN_k$, (\ref{optimization problem1}) is transformed as
\begin{align}
	&\max_{\tilde{\boldsymbol{\theta}}}~~2Re\{\gamma\tilde{\boldsymbol{\theta}}^H\mathbf{q}^t\}\nonumber\\
	&~\text{s.t.}~~~|\tilde{\theta_s}|=1,
\end{align}
where
\begin{align}
	\gamma
	&=\sqrt{\frac{P_I}{\|\tilde{\mathbf{\Theta}}\mathbf{H}\mathbf{V}\mathbf{s}\|^2+\|\tilde{\mathbf{\Theta}}\|_F^2\sigma_n^2}},
\end{align}
and $\tilde{\theta_s}$ is the $s$-th element of $\tilde{\boldsymbol{\theta}}$, $\tilde{\mathbf{\Theta}}=\text{diag}(\tilde{\boldsymbol{\theta}})$ and $\mathbf{q}^t=(\lambda_{max}\mathbf{I}-\mathbf{\Omega})\gamma\tilde{\boldsymbol{\theta}}^t-\mathbf{c}^*$. Then the optimal solution of (\ref{optimization problem1}) is shown as 
\begin{align}\label{MM solution}
	\boldsymbol{\theta}^{t+1}
	&=\gamma e^{jarg(\mathbf{q}^t)},
\end{align}
For a clearer presentation of MM algorithm, its specific procedure is listed in the following Algorithm 1.

\begin{algorithm}[htb]  
	\caption{MM Algorithm}  
	\label{MM Algorithm}  
	\begin{algorithmic}[1]    
		\State Initial $t$=1, $\tilde{\boldsymbol{\theta}}^0$, error tolerance $\epsilon=10^{-6}$. And acquire the value of $f(\boldsymbol{\theta}^1)$ in (\ref{optimization problem1}); 
		\label{MM1}
		\State Calculate $\mathbf{q}^t=(\lambda_{max}\mathbf{I}-\mathbf{\Omega})\gamma\tilde{\boldsymbol{\theta}}^t-\mathbf{c}^*$ and Update $\boldsymbol{\theta}^{t+1}$ based on (\ref{MM solution}); 
		\label{MM2}   
		\State Calculate  $f(\boldsymbol{\theta}^{t+1})$, if $|f(\boldsymbol{\theta}^{t+1})-f(\boldsymbol{\theta}^{t})|/f(\boldsymbol{\theta}^{t+1})\leq\epsilon$ holds, end loop; If not, back to step 2.  
		\label{MM4}  
	\end{algorithmic}  
\end{algorithm}

The details of entire algorithm about proposed WMMSE-PC is shown in Algorithm 2. Noticed that the objective function of (\ref{the optimization problem}) is limited to the finite value under the power constraints, which implied that the Algorithm 2 can ensure convergence. 

\begin{algorithm}[htb]  
	\caption{The proposed WMMSE-PC Algorithm}  
	\label{total Algorithm}  
	\begin{algorithmic}[1]    
		\State Initial $s$=1, the maximum number of iterations $s_{max}=200$, error tolerance $\epsilon$. Input the initialization $\mathbf{V}^{(1)}$, $\mathbf{\Theta}^{(1)}$, calculate the value of  (\ref{the optimization problem}) and represented as $Obj(\mathbf{V}^{(1)}$, $\mathbf{\Theta}^{(1)})$; 
		\label{total1}
		\State Fixed $\mathbf{V}^{(s)}$, $\mathbf{\Theta}^{(s)}$, obtain  $\mathbf{U}^{(s)}$ in (\ref{U_opt});  
		\label{total2}  
		\State  Fixed $\mathbf{V}^{(s)}$, $\mathbf{\Theta}^{(s)}$ and $\mathbf{U}^{(s)}$, calculate $\mathbf{W}^{(s)}$ in (\ref{W});  
		\label{total3}  
		\State Fixed $\mathbf{W}^{(s)}$, $\mathbf{\Theta}^{(s)}$ and $\mathbf{U}^{(s)}$, obtain  $\mathbf{V}^{(s+1)}$ by solving problem (\ref{V_opt});  
		\label{total4}
		\State Fixed $\mathbf{W}^{(s)}$, $\mathbf{U}^{(s)}$ and $\mathbf{V}^{(s+1)}$, calculate  $\mathbf{\Theta}^{(s+1)}$ by working out problem (\ref{Theta_opt});
		\label{total5}
		\State if $s\geq s_{max}$ or $|Obj(\mathbf{V}^{(s+1)}$, $\mathbf{\Theta}^{(s+1)})-Obj(\mathbf{V}^{(s)}$, $\mathbf{\Theta}^{(s)})|/Obj(\mathbf{V}^{(s)}$, $\mathbf{\Theta}^{(s)})\textless\epsilon$, end loop. If not, back to step 2.
		\label{total6}
	\end{algorithmic}  
\end{algorithm}

\subsection{PROPOSED MAX-TR-SVD}

Based on (\ref{the optimization problem}), the overall optimization problem in proposed Max-TR-SVD can be expressed as
\begin{align}\label{the optimization problem2}
	&\max_{\mathbf{\Theta},\mathbf{V}}~~\mathbf{G}^H\mathbf{\Theta}\mathbf{H}\mathbf{V}\mathbf{s}\mathbf{s}^H\mathbf{V}^H\mathbf{H}^H\mathbf{\Theta}^H\mathbf{G}\nonumber\\
	&~~~\text{s.t.}~~\|\mathbf{V}\|_F^2\leq P_B,\nonumber\\
	&~~~~~~~~\|\mathbf{\Theta}\mathbf{H}\mathbf{V}\mathbf{s}\|^2+\|\mathbf{\Theta}\|_F^2\sigma_n^2\leq P_I.		
\end{align}

According to SVD theorem
\begin{align}
	\mathbf{H}
	&=\mathbf{U}_1\mathbf{\Sigma}_1\mathbf{V}_1^H,
\end{align}
and
\begin{align}
	\mathbf{G}^H
	&=\mathbf{U}_2\mathbf{\Sigma}_2\mathbf{V}_2^H,
\end{align}
then the transmit beamforming $\mathbf{V}$ and receive beamforming $\mathbf{U}$ are obtained from $\mathbf{V}_1$ and $\mathbf{U}_2$, respectively.

(\ref{the optimization problem2}) can be converted to 
\begin{align}\label{theta45}
	&\max_{\mathbf{\Theta}}~~tr(\mathbf{G}^H\mathbf{\Theta}\mathbf{H}\mathbf{V}\mathbf{s}\mathbf{s}^H\mathbf{V}^H\mathbf{H}^H\mathbf{\Theta}^H\mathbf{G})\nonumber\\
	&~~\text{s.t.}~~\|\mathbf{\Theta}\mathbf{H}\mathbf{V}\mathbf{s}\|^2+\|\mathbf{\Theta}\|_F^2\sigma_n^2\leq P_I.		
\end{align}

Define $\mathbf{\Theta}=\text{diag}(\boldsymbol{\theta})$, (\ref{theta45}) can be rewritten as
\begin{align}\label{the optimization problem3}
	&\max_{\boldsymbol{\theta}}~~\boldsymbol{\theta}^H\text{diag}(\mathbf{H}\mathbf{V}\mathbf{s})^H\mathbf{G}\mathbf{G}^H\text{diag}(\mathbf{H}\mathbf{V}\mathbf{s})\boldsymbol{\theta}\nonumber\\
	&~~\text{s.t.}~~\boldsymbol{\theta}^H(\text{diag}(\mathbf{H}\mathbf{V}\mathbf{s})^H\text{diag}(\mathbf{H}\mathbf{V}\mathbf{s})+\sigma_n^2\mathbf{I})\boldsymbol{\theta}\leq P_I.		
\end{align}

Let us define $P_I=\boldsymbol{\theta}^H(\text{diag}(\mathbf{H}\mathbf{V}\mathbf{s})^H\text{diag}(\mathbf{H}\mathbf{V}\mathbf{s})+\sigma_n^2\mathbf{I})\boldsymbol{\theta}$, and $\boldsymbol{\theta}=\hat{\rho}\hat{\boldsymbol{\theta}}$, where $\hat{\rho}=\|\boldsymbol{\theta}\|$, $\hat{\boldsymbol{\theta}}^H\hat{\boldsymbol{\theta}}=1$, then the (\ref{the optimization problem3}) can be rewritten as
\begin{align}
	&\max_{\hat{\boldsymbol{\theta}}}~~\frac{\hat{\boldsymbol{\theta}}^H\text{diag}(\mathbf{H}\mathbf{V}\mathbf{s})^H\mathbf{G}\mathbf{G}^H\text{diag}(\mathbf{H}\mathbf{V}\mathbf{s})\hat{\boldsymbol{\theta}}}{\hat{\boldsymbol{\theta}}^H\frac{1}{P_I}(\text{diag}(\mathbf{H}\mathbf{V}\mathbf{s})^H\text{diag}(\mathbf{H}\mathbf{V}\mathbf{s})+\sigma_n^2\mathbf{I})\hat{\boldsymbol{\theta}}},
\end{align}
where $\hat{\boldsymbol{\theta}}$ is the eigenvector corresponding to the maximum eigenvalue of $	\hat{\mathbf{D}}$,
where

\begin{align}
	\hat{\mathbf{D}}
	&=\frac{\text{diag}(\mathbf{H}\mathbf{V}\mathbf{s})^H\mathbf{G}\mathbf{G}^H\text{diag}(\mathbf{H}\mathbf{V}\mathbf{s})}{\frac{1}{P_I}(\text{diag}(\mathbf{H}\mathbf{V}\mathbf{s})^H\text{diag}(\mathbf{H}\mathbf{V}\mathbf{s})+\sigma_n^2\mathbf{I})}.
\end{align}

Then
\begin{align}
	\hat{\rho}
	&=\sqrt{\frac{P_I}{\hat{\boldsymbol{\theta}}^H(\text{diag}(\mathbf{H}\mathbf{V}\mathbf{s})^H\text{diag}(\mathbf{H}\mathbf{V}\mathbf{s})+\sigma_n^2\mathbf{I})\hat{\boldsymbol{\theta}}}},
\end{align}
then the $\boldsymbol{\theta}=\hat{\rho}\hat{\boldsymbol{\theta}}$, and $\mathbf{\Theta}=\text{diag}(\boldsymbol{\theta})$.

\section{PERFORMANCE AND COMPLEXITY ANALYSIS}
\subsection{When $P_I$ $\rightarrow$ $\infty$}
In this section, when $P_B$ is fixed, while increasing $P_I$, we  make an analysis about the performance.

Based on (\ref{SINR}), the SINR of $k$-th data-stream is 
\begin{align}\label{jixian}
	&\gamma_k
	=\\&\notag\frac{\frac{P_B}{K}P_{I_k}A_1}{\frac{P_B}{K}\displaystyle\sum_{j=1,j\neq k}^{K}P_{I_j}A_2+P_{I_k}A_3+\displaystyle\sum_{j=1,j\neq k}^{K}P_{I_j}A_4+\sigma_z^2\|\mathbf{u}_k^H\|^2},
\end{align}
where
\begin{align}
	A_1
	&=\frac{\|\mathbf{u}_k^H\mathbf{G}_k^H\|^2\|\text{diag}(\tilde{\boldsymbol{\theta}}_k)\|^2\|\mathbf{H}_{k}\mathbf{v}_ks_k\|^2}{\|\text{diag}(\mathbf{H}_{k}\mathbf{v}_k)	\tilde{\boldsymbol{\theta}}_k\|^2+\sigma_k^2},
\end{align}
\begin{align}
	A_2
	&=\frac{\|\mathbf{u}_k^H\mathbf{G}_j^H\|^2\|\text{diag}(\tilde{\boldsymbol{\theta}}_j)\|^2\|\mathbf{H}_{j}\mathbf{v}_js_j\|^2}{\|\text{diag}(\mathbf{H}_{j}\mathbf{v}_j)	\tilde{\boldsymbol{\theta}}_j\|^2+\sigma_j^2},
\end{align}
\begin{align}
	A_3
	&=\frac{\sigma_k^2\|\mathbf{u}_k^H\mathbf{G}_k^H\|^2\|\text{diag}(\tilde{\boldsymbol{\theta}}_k)\|^2}{\|\text{diag}(\mathbf{H}_{k}\mathbf{v}_k)	\tilde{\boldsymbol{\theta}}_k\|^2+\sigma_k^2},
\end{align}
\begin{align}
	A_4
	&=\frac{\sigma_j^2\|\mathbf{u}_k^H\mathbf{G}_j^H\|^2\|\text{diag}(\tilde{\boldsymbol{\theta}}_j)\|^2}{\|\text{diag}(\mathbf{H}_{j}\mathbf{v}_j)	\tilde{\boldsymbol{\theta}}_j\|^2+\sigma_j^2}.
\end{align}

Noted that when $i\neq k$, the 
$\mathbf{u}_i^H\mathbf{G}_k^H$ $\rightarrow 0$, we have $A_2$ and $A_4$ are $\rightarrow 0$. At this point, (\ref{jixian}) can be rewritten as
\begin{align}
	\gamma_k
	&=\frac{\frac{P_B}{K}P_{I_k}A_1}{P_{I_k}A_3+\sigma_z^2\|\mathbf{u}_k^H\|^2}.
\end{align}

When $P_I$ $\rightarrow$ $\infty$ and all other elements are fixed, then $P_{I_k}$ $\rightarrow$ $\infty$, finally we have
\begin{align}\label{huajianhou}
	\gamma_k
	&=\frac{P_BA_1}{KA_3}.
\end{align}
We can find that it is a constant, which means the achievable rate will reach to a platform period with a continuous increases in the value of $P_I$.

Similar to the derivation of (\ref{huajianhou}), we can convert (\ref{hesulv}) to 
\begin{align}\label{zhengkuaihuajian}
	R_{m2}
	&=\log_2|\mathbf{I}_{K}+\frac{P_I\mathbf{Q_1}}{P_I\mathbf{Q_2}+\sigma^{2}_{z}\mathbf{U}^H\mathbf{U}}|,
\end{align}
where
\begin{align}
	\mathbf{Q_1}
	&=\frac{\mathbf{U}^H\mathbf{G}^H\tilde{\mathbf{\Theta}}\mathbf{H}\mathbf{V}\mathbf{s}\mathbf{s}^H\mathbf{V}^H\mathbf{H}^H\tilde{\mathbf{\Theta}}^H\mathbf{G}\mathbf{U}}{\|\tilde{\mathbf{\Theta}}\mathbf{H}\mathbf{V}\mathbf{s}\|^2+\|\tilde{\mathbf{\Theta}}\|_F^2\sigma_n^2},
\end{align}
\begin{align}
	\mathbf{Q_2}
	&=\frac{\sigma^{2}_{n}\mathbf{U}^H\mathbf{G}^H\tilde{\mathbf{\Theta}}\tilde{\mathbf{\Theta}}^H\mathbf{G}\mathbf{U}}{\|\tilde{\mathbf{\Theta}}\mathbf{H}\mathbf{V}\mathbf{s}\|^2+\|\tilde{\mathbf{\Theta}}\|_F^2\sigma_n^2}.
\end{align}

When $P_I$ $\rightarrow$ $\infty$ and all other elements are fixed, (\ref{zhengkuaihuajian}) reduces to
\begin{align}\label{zhengtihuajianhou}
	R_{m2}
	&=\log_2|\mathbf{I}_{K}+\frac{\mathbf{Q_1}}{\mathbf{Q_2}}|,
\end{align}
which is also a constant.

\subsection{The relationship between $\gamma_0$ and $\gamma_u$}

In this section, the impact of average SINR at the $k$-th active IRS on the $k$ data-stream receive SINR at the user will be shown.

The receive signal about the $n$-th element at IRS $k$ is recast as
\begin{align}\label{yn}
	y_n
	&=\mathbf{h}^H_k(n)\displaystyle \sum_{j=1}^{K}\mathbf{v}_js_j+w_i(n),
\end{align}
where $\mathbf{h}_k(n)\in\mathbb{C}^{M\times 1}$ denotes the transmission channel from the BS to the $n$-th component element of  IRS $k$. $w_i(n)$ is the AWGN at the $n$-th component element of  IRS $k$ and $w_i(n)\sim \mathcal{CN}\left({0,{~\sigma^{2}_{i}}}\right)$.

Since $k\neq j$,  $\mathbf{H}_k\mathbf{v}_{j}=\mathbf{0}$, then $\mathbf{h}^H_k(n)\mathbf{v}_{j}=0$,  (\ref{yn}) can be converted as
\begin{align}
	y_n
	&=\mathbf{h}^H_k(n)\mathbf{v}_ks_k+w_i(n).
\end{align}

Then the average SINR at $k$-th IRS is 
\begin{align}\label{gamma_0}
	\gamma_0
	&=\frac{P_B\displaystyle\sum_{n=1}^{N_k}|\mathbf{h}^H_k(n)\mathbf{v}_k|^2}{N_k\sigma_i^2}.
\end{align}

The $k$-th data stream receive signal based on (\ref{yk}) at user may be expressed in the vector form 
\begin{align}\label{yu}
	y_u
	&=\displaystyle \sum_{n=1}^{N_k}\mathbf{u}_k^H\mathbf{g}_k(n)\varphi_k(n)\mathbf{h}^H_{k}(n)\mathbf{v}_ks_k\notag\\&+\displaystyle \sum_{n=1}^{N_k}\mathbf{u}_k^H\mathbf{g}_k(n)\varphi_k(n)w_{i}(n)\notag\\&+\displaystyle \sum_{n=1}^{N_k}\displaystyle \sum_{j=1,j \neq k}^{K}\mathbf{u}_k^H\mathbf{g}_j(n)\varphi_j(n)\mathbf{h}^H_{j}(n)\mathbf{v}_js_j\notag\\&+\displaystyle \sum_{n=1}^{N_k}\displaystyle \sum_{j=1,j\neq k}^{K}\mathbf{u}_k^H\mathbf{g}_j(n)\varphi_j(n)w_{i}(n)+\mathbf{u}_k^H\mathbf{z},
\end{align}
where $\mathbf{g}_k(n)\in\mathbb{C}^{N_u\times 1}$ stands for the channel from the $n$-th element of  IRS $k$ to user, $\varphi_k(n)$ is the $n$-th element of $\mathbf{\Theta}_k$. Similarly, when $i\neq k$,  
$\mathbf{u}_i^H\mathbf{G}_k^H$ $\rightarrow$ $\mathbf{0}$, then $\mathbf{u}_i^H\mathbf{g}_j(n)$ $\rightarrow$ 0, then (\ref{yu}) is simplified as
\begin{align}
	y_u
	&=\displaystyle \sum_{n=1}^{N_k}\mathbf{u}_k^H\mathbf{g}_k(n)\varphi_k(n)\mathbf{h}^H_{k}(n)\mathbf{v}_ks_k\notag\\&+\displaystyle \sum_{n=1}^{N_k}\mathbf{u}_k^H\mathbf{g}_k(n)\varphi_k(n)w_{i}(n)+\mathbf{u}_k^H\mathbf{z}.
\end{align}

Then we can acquire the power of the $k$-th data-stream receive signal at user is
\begin{align}
	\mathbb{E}\{y_u^Hy_u\}
	&=\displaystyle \sum_{n=1}^{N_k}|\mathbf{u}_k^H\mathbf{g}_k(n)\varphi_k(n)\mathbf{h}^H_{k}(n)\mathbf{v}_k|^2\notag\\&+\displaystyle \sum_{n=1}^{N_k}\sigma_i^2|\mathbf{u}_k^H\mathbf{g}_k(n)\varphi_k(n)|^2+\sigma_z^2\|\mathbf{u}_k^H\|^2.
\end{align}

The signal reflected through the $n$-th reflect element of $k$-th small IRS is
\begin{align}
	y_i(n)
	&=\varphi_k(n)\mathbf{h}^H_k(n)\mathbf{v}_ks_k+\varphi_k(n)w_i(n).
\end{align}

The average total power of all $N_k$ elements of $k$-th small IRS is
\begin{align}
	P_{I_k}
	&=P_B\displaystyle \sum_{n=1}^{N_k}|\varphi_k(n)\mathbf{h}^H_k(n)\mathbf{v}_k|^2+\sigma_i^2\displaystyle \sum_{n=1}^{N_k}|\varphi_k(n)|^2.
\end{align}

Let us define
\begin{align}\label{vpk}
	|\varphi_k(n)|
	&=\lambda_k=\sqrt{\frac{P_{I_k}}{P_B\displaystyle \sum_{n=1}^{N_k}|\mathbf{h}^H_k(n)\mathbf{v}_k|^2+\sigma_i^2N_k}}.
\end{align}

The SINR of $k$-th data stream receive signal at user is
\begin{align}\label{gammau}
	\gamma_u
	&=\frac{P_B\displaystyle \sum_{n=1}^{N_k}|\mathbf{u}_k^H\mathbf{g}_k(n)\varphi_k(n)\mathbf{h}^H_{k}(n)\mathbf{v}_k|^2}{\displaystyle \sum_{n=1}^{N_k}\sigma_i^2|\mathbf{u}_k^H\mathbf{g}_k(n)\varphi_k(n)|^2+\sigma_z^2\|\mathbf{u}_k^H\|^2}.
\end{align}

Then substituting (\ref{vpk}) in (\ref{gammau}), and when $P_I$ $\rightarrow$ $\infty$, we can transform (\ref{gammau}) to 
\begin{align}\label{gamma_u}
	\gamma_u
	&=\frac{P_B\displaystyle \sum_{n=1}^{N_k}|\mathbf{u}_k^H\mathbf{g}_k(n)\mathbf{h}^H_{k}(n)\mathbf{v}_k|^2}{\sigma_i^2\displaystyle \sum_{n=1}^{N_k}|\mathbf{u}_k^H\mathbf{g}_k(n)|^2}.
\end{align}

According to (\ref{gamma_0}) and (\ref{gamma_u}), the relationship between the $\gamma_u$  and the $\gamma_0$ is shown as 
\begin{align}
	\gamma_u
	&=C_1\gamma_0,
\end{align}
where 
\begin{align}
	C_1
	&=\frac{E[|\mathbf{u}_k^H\mathbf{g}_k(n)\mathbf{h}^H_{k}(n)\mathbf{v}_k|^2]}{E[|\mathbf{u}_k^H\mathbf{g}_k(n)|^2]E[|\mathbf{h}^H_k(n)\mathbf{v}_k|^2]}.
\end{align}
It can be found that the receive SINR at user of $k$-th data-stream is proportional to $\gamma_0$ when $N_k$ is fixed. When reducing the noise at IRS elements or increasing the transmit power $P_B$, the receive SINR at user grows linearly.

The complexity of proposed three methods are shown as follows. Their complexity is expressed as a growing order (floating point operations per second (FLOPs)): Max-TR-SVD ($\mathcal{O}(K^3N_k^3+N_uK^2N_k^2+KM^2N_k)$ FLOPs), NSP-ZF-PA ($\mathcal{O}(K(2K^3N_k^3+N_k^2))$ FLOPs), WMMSE-PC ($\mathcal{O}(L_1(K^{3.5}N_k^{3.5}+L_2K^3N_k^3+2K^3))$ FLOPs), where $L_1$ and $L_2$ denote the iteration numbers of WMMSE-PC and MM, respectively.

\section{SIMULATION RESULTS}
In what follows, simulation are presented to assess the convergence and the sum-rate of the proposed three methods. The parameters of simulation are set as: $\sigma_k^2=\sigma_z^2=-40$ dBm, $P_B=30$ dBm, $P_I=0.04$ W. $M=N_u=8$. The coefficient of path loss is chosen as: $l_0=\alpha/d_0^c$, where $c=2$, $\alpha=10^{-2}$ and $d_0$ represents the reference distance. The BS and user are at the original point (0m, 0m) and (100m, 0m), respectively. When there is one large-scale IRS, it is located at (80m, 20m); when there are two small IRSs, it is located at (80m, 20m) and (90m, 30m); when there are four small IRSs, it is located at (80m, 20m), (90m, 30m), (80m, -20m) and (90m, -30m).

\begin{figure}
	\centering
	\includegraphics[width=3.5in]{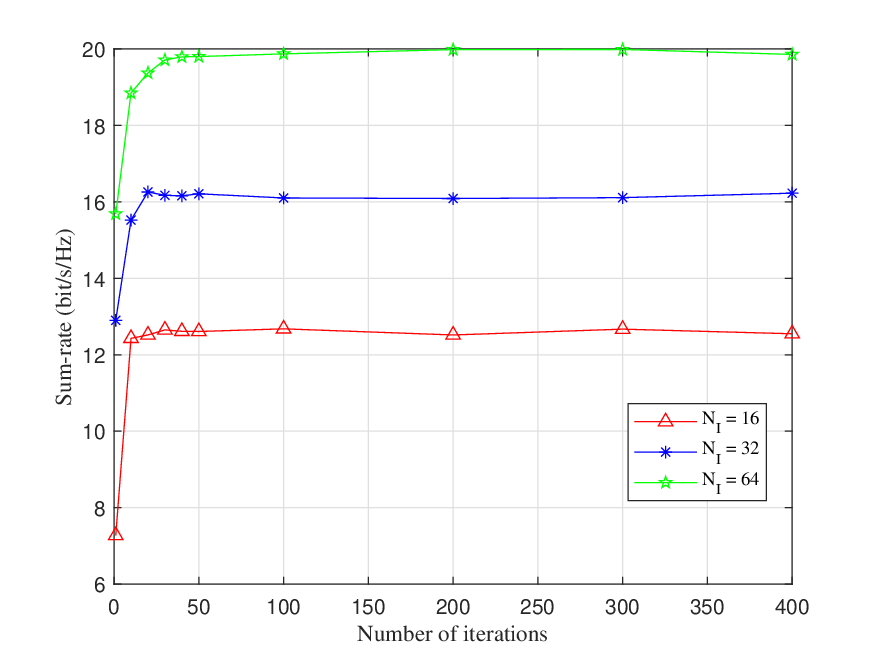}
	\caption{Convergence behavior of WMMSE-PC.}
	\label{fig2} 
\end{figure}

Fig. \ref{fig2} demonstrates the convergence behavior of the proposed WMMSE-PC. It is clearly found that the sum-rate converges fast within fifty iterations with four small IRSs, where the number of the total IRS elements $N_I=$ 16, 32 and 64.  The achieved rate increases  with $N_I$. Additionally, the WMMSE-PC has a similar convergence speed regardless of the value of $N_I$.

\begin{figure}
	\centering
	\includegraphics[width=3.5in]{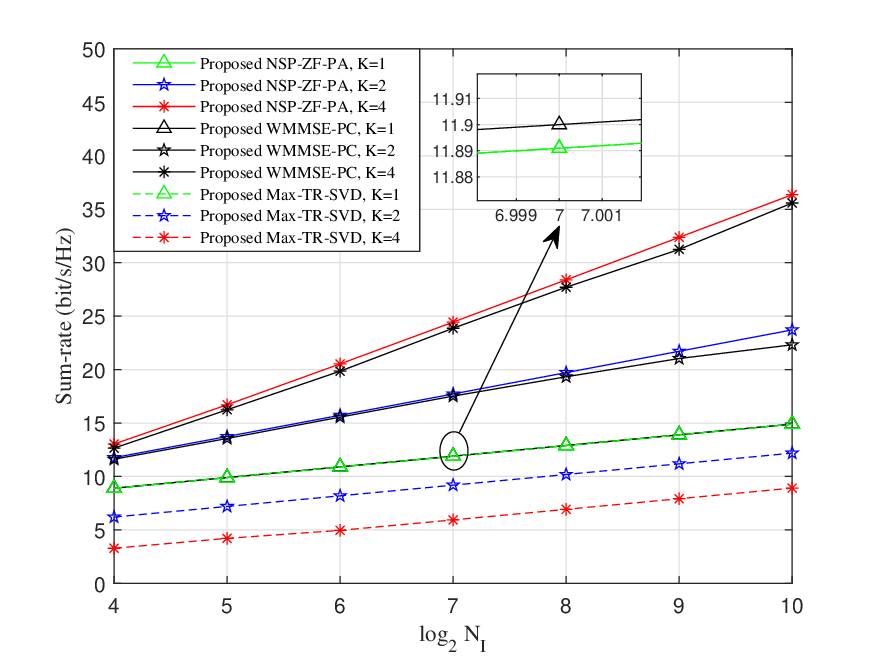} 
	\caption{Sum-rate versus $N_I$.}
	\label{fig3} 
\end{figure}

Fig. \ref{fig3} depicts the sum-rate versus the entire number of active IRS reflecting elements $N_I$ for the three different methods proposed by us. The proposed NSP-ZF-PA and WMMSE-PC have achieved significant rate enhancement over Max-TR-SVD. More importantly, when $K=2$ and $K=4$, the rates of the proposed NSP-ZF-PA and WMMSE-PC are about 1.6 and 2.4 times that with $K=1$ when $N_I=1024$. That is mainly due to the fact that the maximum DoF increases when there are more IRSs. According to (\ref{dof1}), the maximum DoF is four when $K=4$, and there were four data-streams transmitted simultaneously. Thus, we make a conclusion that increasing the number of IRSs with a fixed total number of all IRS elements will have a dominant impact on the performance of rate. This will be further confirmed in the next figure. In other words, a more IRSs means a larger DoF and a higher rate.

\begin{figure}
	\centering
	\includegraphics[width=3.5in]{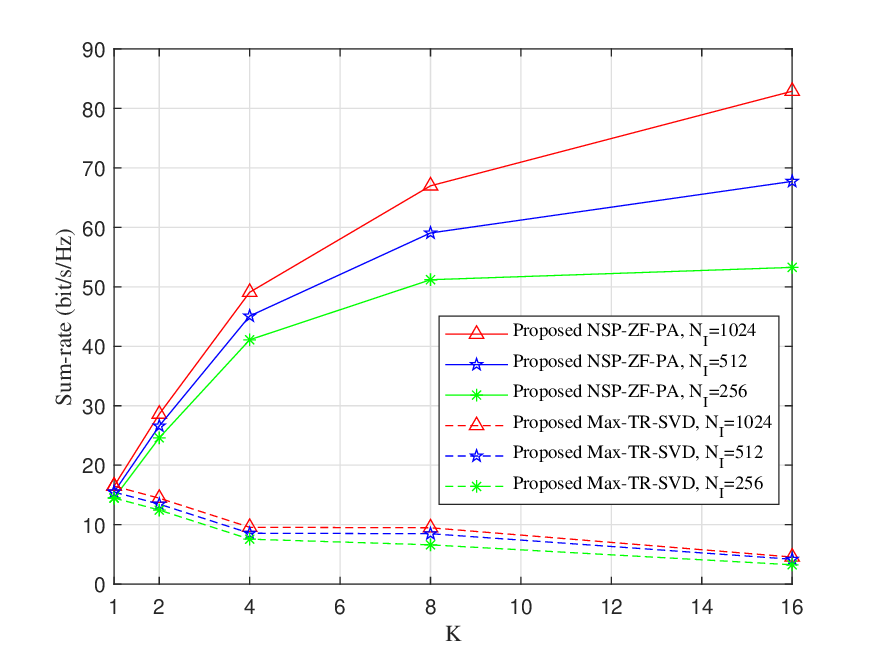} 
	\caption{Sum-rate versus number of IRSs $K$.}
	\label{fig4} 
\end{figure}

Fig. \ref{fig4} demonstrates the sum-rate versus $K$, which is total number of IRSs, where $M=N_u=24$. As the value of $K$ ranges from 1, 2, 4, 8 and 16, the rate of NSP-ZF-PA with sixteen distributed IRSs is about 5 times that of NSP-ZF-PA  with combining all small IRSs as a single large IRS when $N_I=1024$, and about 4.4 times when $N_I=512$, and about 3.7 times when $N_I=256$. This is mainly due to the fact that the  DoF of network increases from 1 to 16. The transmission data-streams also increases from 1 to 16, and it can be transmitted independence among the different data-stream. It can achieve a significant enhancement in this case based on (\ref{Rm1}).

\begin{figure}
	\centering
	\includegraphics[width=3.5in]{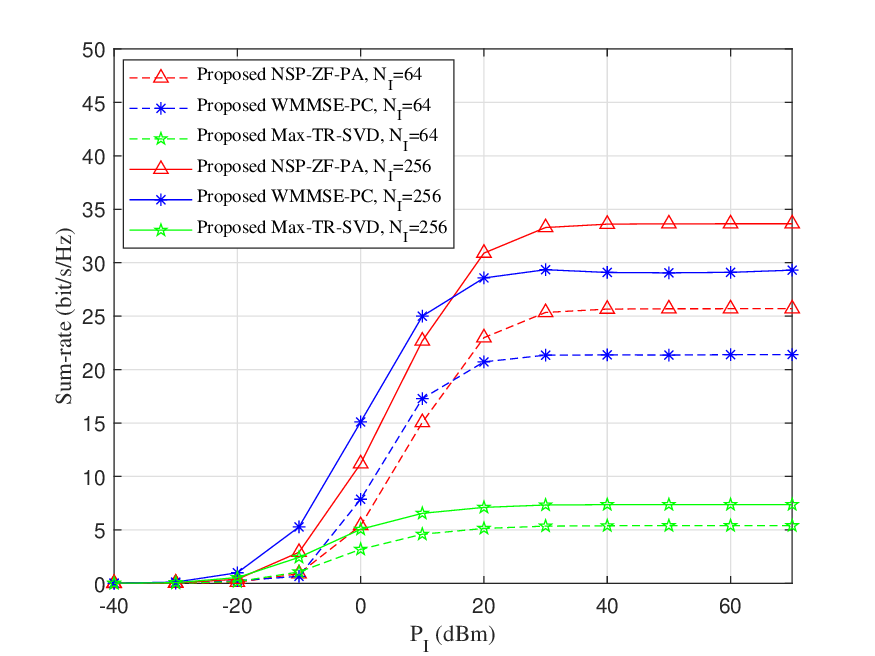} 
	\caption{Sum-rate versus $P_I$.}
	\label{fig5} 
\end{figure}

Fig. \ref{fig5} demonstrates the sum-rate versus the sum power $P_I$ at IRS for the proposed three methods given transmit power $P_B=30$ dBm at BS, where the sum number of IRS $N_I=64, 256$, $K=4$. It can be seen that the proposed NSP-ZF-PA and WMMSE-PC first grow as $P_I\leq20$ dBm increases, but finally reaches a rate ceil  when $P_I\geq20$ dBm. When $P_I$ $\rightarrow$ $\infty$, the achievable rate turns into a constant, which is consistent with the previous conclusions of (\ref{huajianhou}) and (\ref{zhengtihuajianhou}). This result implies that give a fixed transmit power, there is an optimal minimum power budget at IRS to make a highest rate, i.e., the rate upper bound. Placing more power on IRS will waste power. The main reason for this tendency is the effect of noise accumulation at IRS.

\begin{figure}
	\centering
	\includegraphics[width=3.5in]{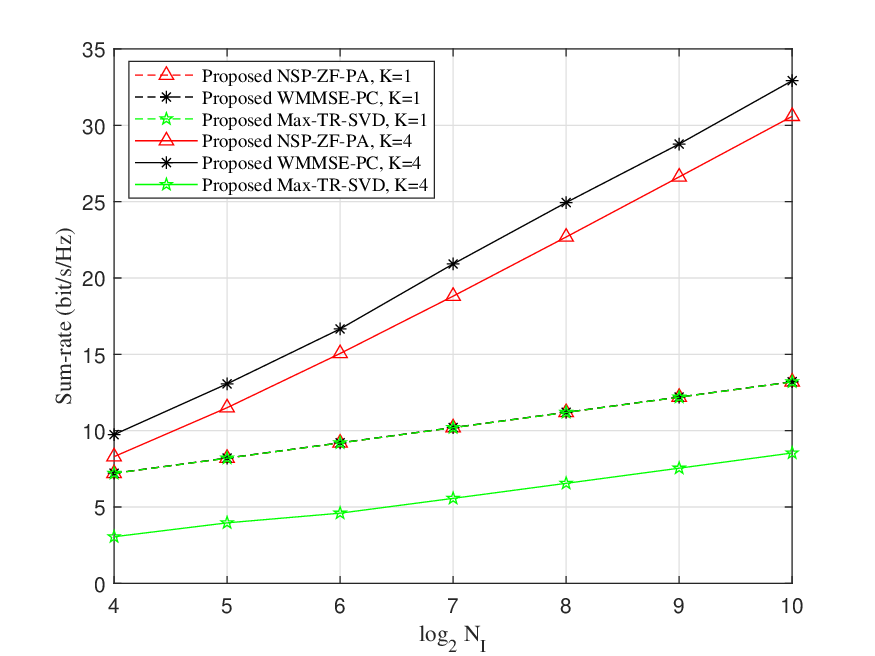} 
	\caption{Achieved sum-rate when $P_I=10$ dBm.}
	\label{fig6} 
\end{figure}

Fig. \ref{fig6} shows the sum-rates of NSP-ZF-PA and WMMSE-PC versus the total number $N_I$ of IRS elements  with $P_I=10$ dBm for two different values of $K$. When $K=4$, the rate performance of WMMSE-PC is better than NSP-ZF-PA. This result is converse to that in Fig. \ref{fig3}. However, the rate growing tendency of both methods is similar to those in the previous figures.

\begin{figure}
	\centering
	\includegraphics[width=3.5in]{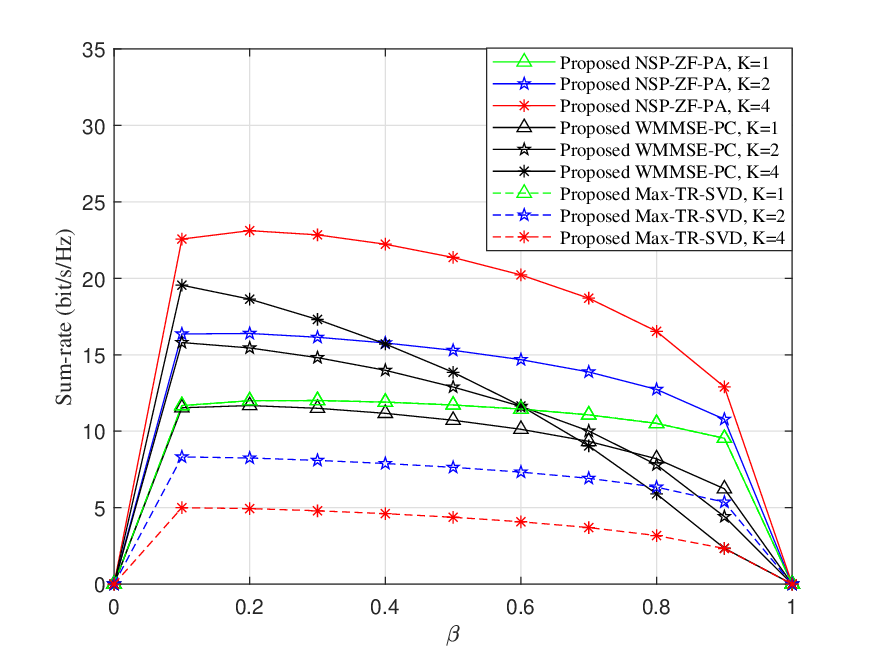} 
	\caption{Sum-rate versus the power allocation factor $\beta$.}
	\label{fig7} 
\end{figure}

Fig. \ref{fig7} depicts the sum-rate versus the PA factor $\beta$ under the total power constraint of $P_T=1$ W, where the power  of all IRSs $P_I=\beta P_T$ , and  $N_I=64$. $\beta=0.1$ means $P_I=0.1$ W. When $K=1$, it can be observed that the optimal PA factor value of $\beta$ of the  proposed NSP-ZF-PA, the WMMSE-PC and Max-TR-SVD approach zero as the number $K$ of IRSs increases. This means less power is allocated to the IRSs.

\begin{figure}
	\centering
	\includegraphics[width=3.5in]{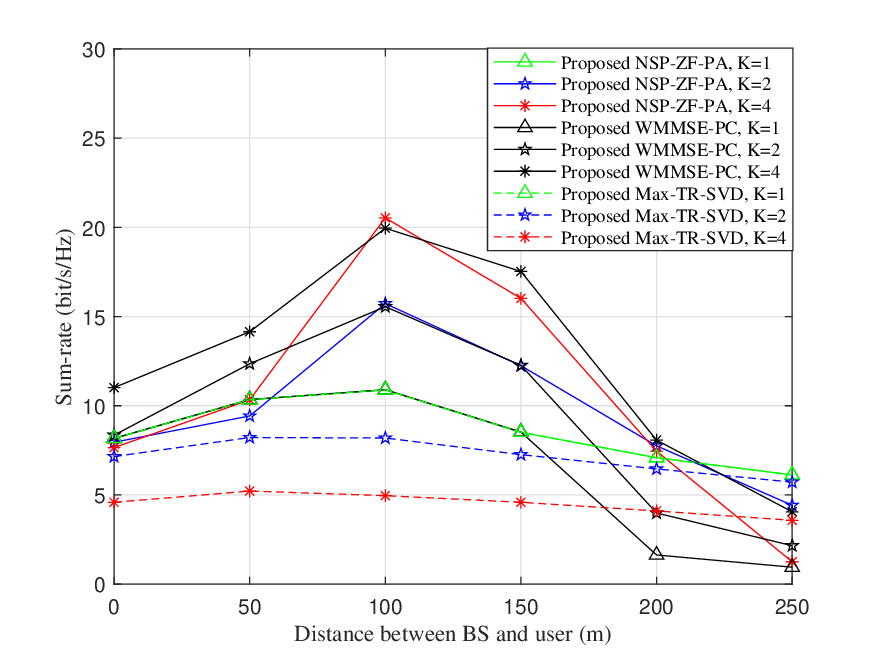} 
	\caption{Sum-rate versus the linear distance between BS and USER.}
	\label{fig8} 
\end{figure}

Fig. \ref{fig8} plots the sum-rate versus the distance between BS and user, where $N_I=64$. It can be shown from Fig. \ref{fig8} that there is a peak value of proposed NSP-ZF-PA and WMMSE-PC when the distance between BS and user is 100m. That is because at this point, the IRS is very  close to user, i.e., the distance from IRS to user is zero,  and can harvest more performance gain. When the distance of BS and user $\geq100$ m, the sum-rates of proposed NSP-ZF-PA and WMMSE-PC reduce as the distance  between BS and user increases, due to the increase in path loss.

\section{CONCLUSION}
In this paper, a novel multi-IRS-aided multi-stream DM network was investigated. This network may form a high $K$ ($\geq 3$) DoFs  to achieve a point-to-point multi-stream by utilizing the distributed small IRSs on UAVs. Subsequently, three high-performance methods, called NSP-ZF-PA, WMMSE-PC, and Max-TR-SVD, were proposed and the computational complexities of these methods were also compared.  Simulation results verified their rates. Particularly, the rate of NSP-ZF-PA with sixteen distributed IRSs was about five times that of NSP-ZF-PA  with combining all small IRSs as a single large IRS. The proposed NSP-ZF-PA and WMMSE-PC performed much better than Max-TR-SVD for multiple-IRS scenario in terms of rate. However, the Max-TR-SVD has the lowest complexity.  And the rate of NSP-ZF-PA  is better than that of WMMSE-PC under the large power budget of IRS, and is worse than that of WMMSE-PC under the low power budget of IRS.

\section*{REFERENCES}

\def\refname{\vadjust{\vspace*{-1em}}} 

\end{document}